\documentclass[aps,twocolumn,showpacs,showkeys,superscriptaddress,unsortedaddress]{revtex4}
\usepackage{graphicx}
\usepackage{amssymb}
\usepackage{bm}
\begin{document}
\setcounter{page}{1}
\title[]{Room-temperature magnetocaloric effect in La$_{0.7}$Sr$_{0.3}$Mn$_{1-x}$\textit{M}$^\prime_x$O$_3$ (\textit{M}$^\prime$ = Al, Ti)}
\author{D. N. H. \surname{Nam}}
\thanks{Corresponding author}
\email{daonhnam@yahoo.com}
\affiliation{Institute of Materials Science, VAST, 18 Hoang-Quoc-Viet, Hanoi, Vietnam}
\affiliation{ICYS, National Institute for Materials Science, Namiki 1-1, Tsukuba, Ibaraki 305-0044, Japan}
\author{N. V. \surname{Dai}}
\author{L. V. \surname{Hong}}
\author{N. X. \surname{Phuc}}
\affiliation{Institute of Materials Science, VAST, 18 Hoang-Quoc-Viet, Hanoi, Vietnam}
\author{S. C. \surname{Yu}}
\affiliation{BK21Physics Program and Department of Physics, Chungbuk National University, Cheongju 361-763, South Korea}
\author{M. \surname{Tachibana}}
\affiliation{ICYS, National Institute for Materials Science, Namiki 1-1, Tsukuba, Ibaraki 305-0044, Japan}
\author{E. \surname{Takayama-Muromachi}}
\affiliation{Advanced Nano Materials Laboratory, National Institute for Materials Science, Namiki 1-1, Tsukuba, Ibaraki 305-0044, Japan}
\date[]{Received \today}

\begin{abstract}
Magnetic entropy and adiabatic temperature changes in and above the room-temperature region has been measured for La$_{0.7}$Sr$_{0.3}$Mn$_{1-x}$\textit{M}$^\prime_x$O$_3$ (\textit{M}$^\prime$ = Al, Ti) by means of magnetization and heat capacity measurements in magnetic fields up to 6 T. The magnetocaloric effect becomes largest at the ferromagnetic ordering temperature $T_\mathrm{c}$ that is tuned to $\sim$300 K by the substitution of Al or Ti for Mn. While the substitution of Al for Mn drastically reduces the entropy change, it extends considerably the working temperature span and improves the relative cooling power. The magnetocaloric effect seems to be only lightly affected by Ti substitution. Although manganites have been considered potential for magnetic refrigerants, the magnetocaloric effect in these materials is limited due to the existence of short-range ferromagnetic correlations above $T_\mathrm{c}$.
\end{abstract}



\maketitle

\section{INTRODUCTION}

The recent invention of a magnetic cooling prototype by Zimm \textit{et al.} \cite{Zimm} opened up a great opportunity for the development of a new generation of refrigerators that are more efficient, inexpensive, and environmentally friendly for replacing the current refrigerators using greenhouse gases that are harmful to environment and contributing to global warming. According to the Maxwell's relations, when the application of a magnetic field ($H_\mathrm{max}$) is adiabatically removed from a ferromagnetic material, cooling occurs due to the change of magnetic entropy,
\begin{equation}
\Delta S_\mathrm{m}\left(\Delta H\right)=-\int_0^{H_\mathrm{max}}\left[\frac{\partial M(T,H)}{\partial T}\right]_{H}dH
\label{eqn1}
\end{equation}
that induces an adiabatic temperature change,
\begin{equation}
\Delta T_\mathrm{a}\left(\Delta H\right)=\int_0^{H_\mathrm{max}}\frac{T}{C_\mathrm{p}(T,H)}\left[\frac{\partial M(T,H)}{\partial T}\right]_{H}dH.
\label{eqn2}
\end{equation}
\begin{figure}[b!]
\includegraphics[width=5.3cm]{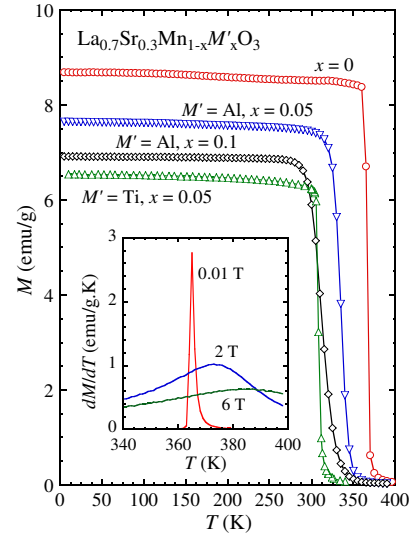}
\caption{Field-cooled $M(T)$ curves for the samples measured in $H=0.01$ T. The inset shows derivative $\partial M/\partial T$ curves of the parent compound, La$_{0.7}$Sr$_{0.3}$MnO$_3$, in $H=0.01$ T, 2 T, and 6 T.} \label{fig.1}
\end{figure}
Here, $C_\mathrm{p}$ is the heat capacity and $\Delta H=H_\mathrm{max}$. The effect is usually largest near the ferromagnetic ordering temperature $T_\mathrm{c}$ where a maximum change of magnetization with temperature is usually obtained. Gd metal was used as a prototypical refrigerant for room-temperature (RT) magnetic cooling \cite{Zimm} since it has a $T_\mathrm{c}=294$ K and a very large magnetocaloric effect (MCE) \cite{Dan'kov,Gschneidner}. The MCE has been found largest in Gd$_5$Ge$_2$Si$_2$ (and related compounds) \cite{Pecharsky,Pecharsky2,Wada}, however this material has a $T_\mathrm{c}$ ($\sim$270 K) somehow far below RT and reaches high magnetocaloric values at room temperature only under large magnetic field changes. Giant RT MCEs have also been observed in MnAs$_{1-x}$Sb$_x$ \cite{Wada}, but unfortunately these materials are not environmentally friendly. The Fe$_{0.49}$Rh$_{0.51}$ alloy exhibits a strong RT MCE comparable to that of Gd but has no practical application because (i) Rh is very expensive and (ii) the MCE is attenuated due to an irreversibility of the antiferromagnetic-ferromagnetic transition \cite{Annaorazov,Annaorazov2}. Perovskite manganites, which are well-known for their colossal magnetoresistance (CMR) effect \cite{Helmolt}, have also been considered potential refrigerants for the future magnetic cooling refrigerators \cite{Gschneidner,Huong}. Although the MCE in manganites is moderate, the materials have certain advantages over the others, such as their low cost, simple fabrication, and easily tunable $T_\mathrm{c}$. A recent study has shown that the substitution of Al and Ti for Mn in La$_{0.7}$Sr$_{0.3}$MnO$_3$ effectively lowers the ferromagnetic ordering temperature $T_\mathrm{c}$ from $\sim$364.5 K to well below 300 K \cite{Nam}. These compounds are therefore suitable for studies of MCE in the room-temperature region. Further, it would be interesting to see how a dilution of the Mn lattice by nonmagnetic elements affects the MCE in the manganite systems.

\begin{figure}[t!]
\includegraphics[width=5.3cm]{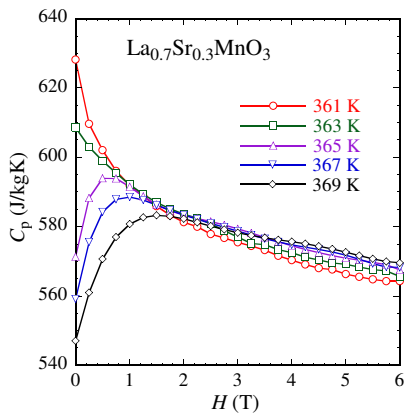}
\caption{Magnetic field dependence of the heat capacity of La$_{0.7}$Sr$_{0.3}$MnO$_3$ measured at different temperatures in the FM-PM phase transition region.} \label{fig.2}
\end{figure}

\begin{figure}[b!]
\includegraphics[width=5.3cm]{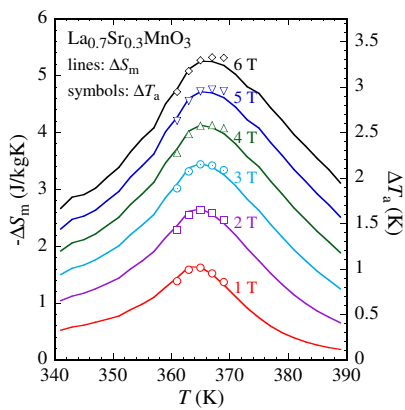}
\caption{MCE of the $x=0$ sample measured with $\Delta H$ varies from 1 T to 6 T. The effect is largest at $T\approx365$ K. Left axis: magnetic entropy change (lines), right axis: adiabatic temperature change (symbols)} \label{fig.3}
\end{figure}

\section{Experiments}
Four samples of La$_{0.7}$Sr$_{0.3}$Mn$_{1-x}M^\prime_x$O$_3$ ($x=0$, 0.05, and 0.1 for $M^\prime=$ Al and $x=0.05$ for $M^\prime=$ Ti) prepared by conventional solid state reaction were pieces taken from the same batches used in a previous work \cite{Nam}. Room-temperature x-ray diffraction measurements show that the samples are essentially single phase of perovskite rhombohedral (space group $R\bar{3}c$) structures. Redox titration experiments (using K$_2$Cr$_2$O$_7$ titrant and C$_{24}$H$_{20}$BaN$_2$O$_6$S$_2$ as the colorimetric indicator) show that the samples are stoichiometric without any significant oxygen excess or deficiency. Magnetic and heat capacity measurements are carried out in a Quantum Design PPMS. The $-\Delta S_\mathrm{m}(T,H)$ and  $\Delta T_\mathrm{a}(T,H)$ values are calculated from the isothermal magnetization $M(H)$ and heat capacity $C_\mathrm{p}(H)$ curves in magnetic fields up to 6 T.

\section{Results and discussion}

As presented in Fig. \ref{fig.1}, the temperature dependent magnetization $M(T)$ curves measured in an applied field of 0.01 T for all the samples show very sharp ferromagnetic (FM)-paramagnetic (PM) phase transitions. The transition temperature $T_\mathrm{c}$ (determined as the temperature where  $\partial M/\partial T$ attains a maximum), which is about 364.5 K for the undoped compound La$_{0.7}$Sr$_{0.3}$MnO$_3$, is effectively tuned to near room temperature by increasing the substitution concentrations. The decrease of $T_\mathrm{c}$ has been explained as a result of the dilution effect that causes a decrease of the molecular field acting on Mn ions according to the molecular field theory \cite{Nam}. The sharp FM-PM phase transitions would promisingly imply a high entropy change at $T_\mathrm{c}$ according to equation \ref{eqn1}. Nevertheless, with increasing magnetic field, the transition becomes strongly broadened, rapidly reducing the $\partial M/\partial T$ value at $T_\mathrm{c}$, as can be seen in the Fig. \ref{fig.1} inset as an example typically presented for the undoped La$_{0.7}$Sr$_{0.3}$MnO$_3$ sample. The broadening of the phase transition with increasing magnetic field consequently inhibits the magnetic entropy change as well as the adiabatic temperature change of the sample.

\begin{figure}[b!]
\includegraphics[width=5.3cm]{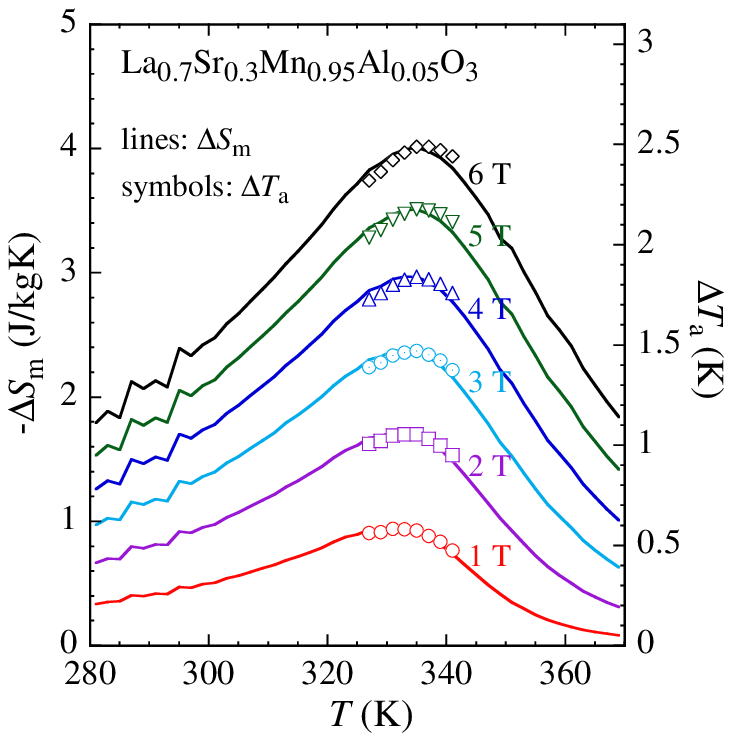}
\caption{MCE of the La$_{0.7}$Sr$_{0.3}$Mn$_{0.95}$Al$_{0.05}$O$_3$ sample with $\Delta H$ varies from 1 T to 6 T. The sample has $T_\mathrm{c}\approx332$ K. Left axis: magnetic entropy change (lines), right axis: adiabatic temperature change (symbols).} \label{fig.4}
\end{figure}

\begin{figure}[t!]
\includegraphics[width=5.3cm]{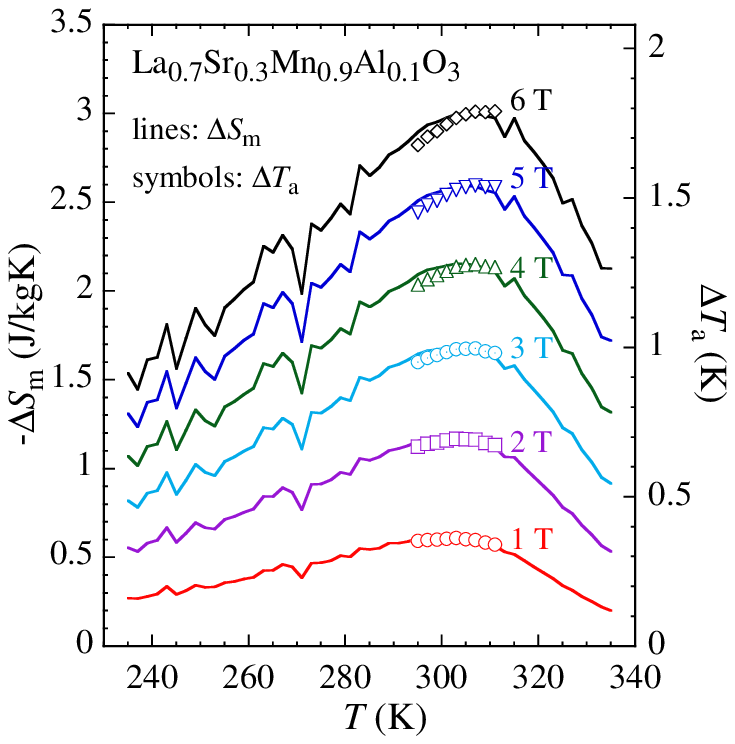}
\caption{MCE of the La$_{0.7}$Sr$_{0.3}$Mn$_{0.9}$Al$_{0.1}$O$_3$ sample with $\Delta H$ varies from 1 T to 6 T. The sample has $T_\mathrm{c}\approx310$ K. Left axis: magnetic entropy change (lines), right axis: adiabatic temperature change (symbols).} \label{fig.5}
\end{figure}

\begin{figure}[t!]
\includegraphics[width=5.3cm]{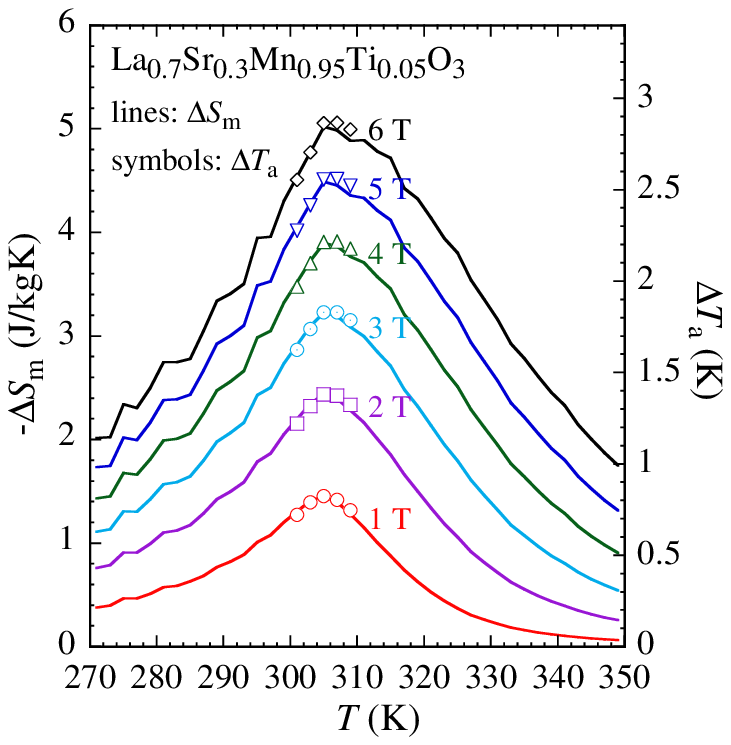}
\caption{MCE of the La$_{0.7}$Sr$_{0.3}$Mn$_{0.95}$Ti$_{0.05}$O$_3$ sample with $\Delta H$ varies from 1 T to 6 T. The sample has $T_\mathrm{c}\approx309$ K. Left axis: magnetic entropy change (lines), right axis: adiabatic temperature change (symbols).} \label{fig.6}
\end{figure}

In order to examine the MCE, isothermal $M(H)$ and heat capacity $C_\mathrm{p}(H)$ measurements are carried out at different temperatures (with temperature interval $\Delta T=2$ K) in the FM-PM transition region for all the samples. The $C_\mathrm{p}(H)$ curves of the undoped sample are typically presented in Fig. \ref{fig.2}. The heat capacity only varies slightly with magnetic field but distinguishably differently between below and above $T_\mathrm{c}$ in the low field region. Above $T_\mathrm{c}$, $C_\mathrm{p}(H)$ increases as the field starts increasing from zero, reaches a maximum and finally decreases at higher fields. At temperatures below $T_\mathrm{c}$, $C_\mathrm{p}(H)$ monotonically decreases with $H$. The $-\Delta S_\mathrm{m}(T,\Delta H)$ and $\Delta T_\mathrm{a}(T,\Delta H)$ data calculated from the $M(H)$ and $C_\mathrm{p}(H)$ curves are plotted in Figs. \ref{fig.3}-\ref{fig.6}. Although the FM-PM phase transition is strongly influenced by magnetic field (Fig. \ref{fig.1} inset), for all the $\Delta H$ values up to 6 T, a maximum of $-\Delta S_\mathrm{m}$ is always observed at about $T_\mathrm{c}$ with only a very small shift to higher temperatures with higher fields. This behavior is due to the fact that, with increasing magnetic field, although the  $\partial M/\partial T$ maximum greatly shifts to higher temperatures, its magnitude is strongly suppressed, leading to the dominance of low field $-\Delta S_\mathrm{m}$ at $T_\mathrm{c}$. The maximum values of $-\Delta S_\mathrm{m}$ and $\Delta T_\mathrm{a}$ of the samples are smaller than that of Gd, but comparable to other manganite systems \cite{Gschneidner,Huong}. For the Al substituted samples, the entropy change decreases considerably with increasing $x$. However, the relative cooling power ($RCP$), $RCP=-\Delta S_\mathrm{m}(T,\Delta H)\times\delta_\mathrm{FWHM}$, where $\delta_\mathrm{FWHM}$ denotes the full width at half maximum, is improved. For examples, with $\Delta H=2$ T, $RCP$ increases from 80.3 J/kg for $x=0$ to 100.3 J/kg for $x=0.05$ and 108.8 J/kg for $x=0.1$. Refrigerants with a wide working temperature span and high $RCP$ are in fact very beneficial to magnetic cooling applications. It is interesting that while $T_\mathrm{c}$ is largely decreased, the Ti substitution only slightly affects the magnetic cooling effect.

\begin{figure}[b!]
\includegraphics[width=5.3cm]{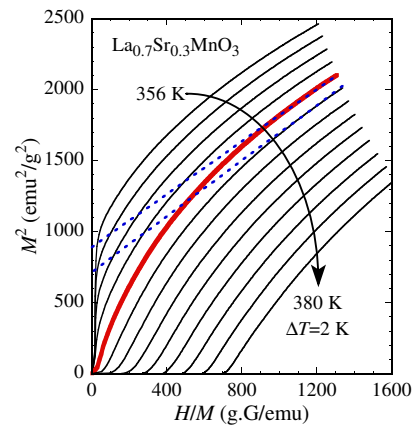}
\caption{Arrott's plots for La$_{0.7}$Sr$_{0.3}$MnO$_3$ carried out in the transition region. The bold curve represents the measurement at 364 K, close to $T_\mathrm{c}=364.5$ K. The dashed lines which present extrapolations of the high-field linear portion of the 364 K and 366 K curves to zero field both intersect the $y$-axis at $M^2>0$ indicating the existence of a field-induced ferromagnetism in the transition region.} \label{fig.7}
\end{figure}

Analyzes from the x-ray diffraction spectra indicate that the unit cell volume $V$ decreases with Al substitution (from 350.98 {\AA}$^3$ for $x=0$ to 349.34 {\AA}$^3$ for $x=0.1$) while it increases with Ti substitution (to 351.91 {\AA}$^3$ for $x=0.05$). Because the ionic size of Al$^{3+}$ (0.535 {\AA}) is smaller than that of Mn$^{3+}$ (0.645 {\AA}), and Ti$^{4+}$ (0.605 {\AA}) is larger than Mn$^{4+}$ (0.530 {\AA}), the changes in the unit cell volume would suggest that Al and Ti are selectively substituted for Mn$^{3+}$ and Mn$^{4+}$, respectively. Since both Al$^{3+}$ and Ti$^{4+}$ do not carry a net spin moment, the substitution thus results in a decrease of the average magnetic moment per transition metal ion, which is theoretically from 3.7 $\mu_\mathrm{B}$ (for $x=0$) to 3.5 $\mu_\mathrm{B}$ ($x=0.05$) and 3.3 $\mu_\mathrm{B}$ ($x=0.1$) for Al substitution and to 3.55 $\mu_\mathrm{B}$ ($x=0.05$) for Ti substitution. The experimental values determined at low temperature and in high field follow the same trend: from 3.65 $\mu_\mathrm{B}$ ($x=0$) to 3.44 $\mu_\mathrm{B}$ ($x=0.05$) and 3.1 $\mu_\mathrm{B}$ ($x=0.1$) for the Al-substituted samples and to 3.54 $\mu_\mathrm{B}$ for the $x=0.05$ Ti-substituted sample. Although $\partial M/\partial T$ is the factor that determines the MCE magnitude according to equations \ref{eqn1} and \ref{eqn2}, the magnetization value itself practically plays an important role in real systems. Since most strength of the magnetization collapses in the phase transition region, the system with a higher spontaneous magnetization may have a higher $\partial M/\partial T$ at $T_\mathrm{c}$. In fact, Gd metal, which has a very high magnetic moment of up to 8 $\mu_\mathrm{B}$/at., and its based compounds are those ferromagnets that exhibit strongest MCEs ever found to date. Our results here clearly show that the MCE in our samples varies consistently with the change of the average magnetic moment, giving an explanation to the weaker suppression of MCE in the case of Ti substitution than in the case of Al substitution.

Comparing to other magnetic cooling materials known to date, the MCE observed in manganites is not very large in general. There have been many convincing evidences that short-range FM correlations, or clusters, persist up to high temperatures well above $T_\mathrm{c}$ in manganites \cite{Teresa,Teresa2,Kapusta,Fernandez-Baca}. The FM clusters grow greatly in size under an influence of magnetic field when approaching $T_\mathrm{c}$ from above. Consequently, the FM-PM transition is quickly broadened, shifting the apparent $T_\mathrm{c}$ to higher temperatures with increasing applied field as illustrated in the inset of Fig. \ref{fig.1}. The evidence for the existence of FM clusters above $T_\mathrm{c}$ in our samples could also be seen in the $M(H)$ curves where a true paramagnetic behavior is not observed even at temperatures far above $T_\mathrm{c}$. The Arrott plots \cite{Arrott,Kouvel}, $M^2$ vs. $H/M$, presented in Fig. \ref{fig.7} for La$_{0.7}$Sr$_{0.3}$MnO$_3$ unambiguously indicate that the system is not in a pure PM state above $T_\mathrm{c}=364.5$ K, but a field-induced ferromagnetic state that diffuses to higher temperatures in higher fields. For instance, above $H=3$ T, induced ferromagnetism is found to exist at temperatures up to 374 K. The broadening of the phase transition due to the field-induced ferromagnetism above $T_\mathrm{c}$ results in the suppression of $\partial M/\partial T$ and therefore inhibits the MCE.
\section{CONCLUSIONS}
The ferromagnetic ordering temperature $T_\mathrm{c}$ of La$_{0.7}$Sr$_{0.3}$Mn$_{1-x}$\textit{M}$^\prime_x$O$_3$ has been tuned to near room temperature by partial substitutions of Al or Ti for Mn. All the samples exhibit a maximum of magnetic entropy and adiabatic temperature changes at the phase transition. For the Al-substituted series, the magnitude of MCE is drastically suppressed; however, the relative cooling power is considerably improved. By contrast, Ti substitution has only a little effect on the MCE. The suppression of the MCE with substitution concentration is well consistent with the decrease of saturation magnetization. Although manganites have been considered to be potential candidates for magnetic refrigeration, the existence of magnetic clusters and field induced ferromagnetism above $T_\mathrm{c}$ in these materials is in fact the major obstacle for reaching large MCE.
\begin{acknowledgments}
A part of this work has been performed using facilities of the State Key Labs (IMS, VAST). N. V. Dai thanks for the support from the BK21 Physics Program at Chungbuk National University. This work was supported by the Special Coordination Funds for Promoting Science and Technology from MEXT, Japan.
\end{acknowledgments}


\begin{references}
\bibitem{Zimm}
    C. Zimm, A. Jastrab, A. Sternberg, J. Cryog. Eng. {\bf 43}, 1759 (1998).
\bibitem{Dan'kov}
    S. Yu. Dan'kov, A. M. Tishin, V. K. Pecharsky, and K. A. Gschneidner Jr., Phys. Rev. B {\bf 57}, 3478 (1998).
\bibitem{Gschneidner}
    K. A. Gschneidner Jr, V. K. Pecharsky, and A. O. Tsokol, Rep. Prog. Phys. {\bf 68}, 1479 (2005).
\bibitem{Pecharsky}
    V. K. Pecharsky, K. A. Gschneidner Jr., J. Appl. Phys. Lett. {\bf 70}, 3299 (1997).
\bibitem{Pecharsky2}
    V. K. Pecharsky and K. A. Gschneidner Jr., Phys. Rev. Lett. {\bf 78}, 4494 (1997).
\bibitem{Wada}
    H. Wada and Y. Tanabe, J. Appl. Phys. Lett. {\bf 79}, 3302 (2001).
\bibitem{Annaorazov}
    M. P. Annaorazov, K. A. Asatryan, G. Myalikgulyev, S. A. Nikitin, A. M. Tishin, and L. A. Tyurin, Cryogenics {\bf 32}, 867 (1992).
\bibitem{Annaorazov2}
    M. P. Annaorazov, S. A. Nikitin, A. L. Tyurin, K. A. Asatryan, and A. Kh. Dovletov, J. Appl. Phys. {\bf 79}, 1689 (1996).
%
\bibitem{Helmolt}
    R. V. Helmolt, J. Wecker, B. Holzapfel, L. Schultz, and K. Samwer, Phys. Rev. Lett. {\bf 71}, 2331 (1993).
\bibitem{Huong}
    Manh-Huong Phan and Seong-Cho Yu, J. Mag. Mag. Mater. {\bf 308}, 325 (2007).
\bibitem{Nam}
    D. N. H. Nam, L. V. Bau, N. V. Khiem, N. V. Dai, L. V. Hong, N. X. Phuc, R. S. Newrock, and P. Nordblad, Phys. Rev. B {\bf 73}, 184430 (2006).
\bibitem{Teresa}
    J. M. De Teresa, M. R. Ibarra, P. A. Algarabel, C. Ritter, C. Marquina, J. Blasco, J. García, A. del Moral, and Z. Arnold, Nature (London) {\bf 386}, 256 (1997).
\bibitem{Teresa2}
    J. M. De Teresa, M. R. Ibarra, J. Blasco, J. García, P. A. Algarabel, Z. Arnold, K. Kamenev, C. Ritter, and R. von Helmolt, Phys. Rev. B {\bf 54}, 1187 (1996).
\bibitem{Kapusta}
    C. Kapusta, R. C. Riedl, W. Kocemba, G. J. Tomka, M. R. Ibarra, J. E. de Teresa, M. Viret, and J. M. D. Coey, J. Phys.: Condens. Matter {\bf 11}, 4079 (1999).
\bibitem{LiuX}
    X. Liu, X. Xu, and Y. Zhang, Phys. Rev. B {\bf 62}, 15 112 (2000).
\bibitem{Fernandez-Baca}
    J. A. Fernandez-Baca, P. Dai, H. Y. Hwang, C. Kloc, and S-W. Cheong, Phys. Rev. Lett. {\bf 80}, 4012 (1998).
\bibitem{Arrott}
    A. Arrott, Phys. Rev. {\bf 108}, 1394 (1957).
\bibitem{Kouvel}
    J. S. Kouvel and M. E. Fisher, Phys. Rev. {\bf 136}, A1626 (1964).
%
\end{references}
\end{document}